\documentstyle[epsfig,12pt]{article}
 
\begin{document}  
\newcommand {\ber} 
{\begin{eqnarray*}} 
 \newcommand {\eer} {\end{eqnarray*}} 
\newcommand {\beq}{\begin{equation}}  
\newcommand {\eeq} {\end{equation}} 
\newcommand {\state} 
[1] {\mid \! \! {#1} \rangle} \newcommand {\eqref} [1] {(\ref {#1})} 
\newcommand{\preprint}[1]{\begin{table}[t] 
           \begin{flushright}               
           \begin{large}{#1}\end{large}     
           \end{flushright}                 
           \end{table}}                     
\def\Acknowledgements{\bigskip  \bigskip {\begin{center} \begin{large} 
             \bf ACKNOWLEDGMENTS \end{large}\end{center}}} 
 
\def\Dslash{\not{\hbox{\kern-4pt $D$}}} 
\def\eqref#1{(\ref{#1})}
\def\cmp#1{{\it Comm. Math. Phys.} {\bf #1}} 
\def\cqg#1{{\it Class. Quantum Grav.} {\bf #1}} 
\def\pl#1{{\it Phys. Lett.} {\bf #1B}} 
\def\prl#1{{\it Phys. Rev. Lett.} {\bf #1}} 
\def\prd#1{{\it Phys. Rev.} {\bf D#1}} 
\def\prr#1{{\it Phys. Rev.} {\bf #1}} 
\def\np#1{{\it Nucl. Phys.} {\bf B#1}} 
\def\ncim#1{{\it Nuovo Cimento} {\bf #1}} 
\def\jmath#1{{\it J. Math. Phys.} {\bf #1}} 
\def\mpl#1{{\it Mod. Phys. Lett.}{\bf A#1}} 
\def\jmp#1{{\it J. Mod. Phys.}{\bf A#1}} 
\def\mycomm#1{\hfill\break{\tt #1}\hfill\break} 
\newcommand{\boldsymbol}[1]{\mbox{\boldmath $#1$}} 

\newcommand{\be}{\begin{equation}}  
\newcommand{\ee}{\end{equation}}  
\newcommand{\eel}[1]{\label{#1}\end{equation}}  
\newcommand{\bea}{\begin{eqnarray}}  
\newcommand{\eea}{\end{eqnarray}}  
\newcommand{\eeal}[1]{\label{#1}\end{eqnarray}}  
\newcommand{\baq}{\begin{equation}\begin{array}{rcl}}  
\newcommand{\eaq}{\end{array}\end{equation}}  
\newcommand{\eaql}[1]{\end{array}\label{#1}\end{equation}}  
\newcommand{\beac}{\begin{equation}\begin{array}{rcl}}  
\newcommand{\eeacn}[1]{\end{array}\label{#1}\end{equation}}  
\newcommand{\ba}{\begin{array}}  
\newcommand{\ea}{\end{array}}  
\newcommand{\non}{\nonumber \\}  
\newcommand{\equ}[1]{(\ref{#1})}  
\newcommand{\al}{\alpha}  
\newcommand{\vp}{\varphi}  
\newcommand{\ga}{\gamma}  
\newcommand{\s}{\sigma}  
\newcommand{\g}{{\cal G}}  
\newcommand{\CR}{\non\cr}  
\begin{titlepage} 
\titlepage 
\rightline{} 
\rightline{WIS/15/04-MAY-DPP} 
\rightline{TAUP 2771-04} 
 
\vskip 1cm 
\centerline{{\Large \bf The spectrum of  states with  one current acting}} 
\centerline{{\Large \bf  
 on the adjoint vacuum of    
massless  ${\rm QCD}_2$ }} 
\vskip 1cm

\centerline{A. Abrashkin$^a$\footnote{arielab@post.tau.ac.il} 
, Y. Frishman$^b$\footnote{yitzhak.frishman@weizmann.ac.il} 
, J. Sonnenschein$^a$\footnote{cobi@post.tau.ac.il}}
 
\begin{center}
 
\em $^a$School of Physics and Astronomy 
\\Beverly and Raymond Sackler Faculty of Exact Sciences 
\\Tel Aviv University, Ramat Aviv, 69978, Israel 
\end{center} 
\begin{center} 
\em $^b$Department of Particle Physics 
\\Weizmann Institute of Science 
\\76100 Rehovot, Israel 
\end{center}

\begin{abstract}  
We consider a ``one current'' state, which is obtained by the application of a   
color current on the ``adjoint'' vacuum. This is done in $QCD_2$, with the   
underlying   
quarks in the fundamental representation. The quarks  are taken to be   
massless, in which case the theory  on  the light-front can be ``currentized'', namely,   
formulated in terms of currents only.  
The adjoint vacuum is shown to be the application of a current derivative,  
at zero momentum, on the singlet vacuum.
We apply the operator $M^2=2P^+P^-$ on these states and find that in general they are not eigenstates of $M^2$ apart from the large $N_f$ limit. 
Problems with infra-red   
regularizations are pointed out.  We discuss the fermionic structure of these states.  
\end{abstract} 
\end{titlepage} 
 
\newpage 
\section{Introduction} 
 
The formulation  of two dimensional massless ${\rm QCD}_2$  in terms of only  colored currents (``currentization'') 
turns out to be very natural once the system is quantized on the light-front. Both the momentum and the  
Hamiltonian, and hence also $M^2$, are expressed in terms of the light-cone colored currents. In fact only  
the left (or right) currents are needed \cite{KS,AS}.  
The currentization was shown to hold for multiflavor fundamental quarks, and adjoint quarks \cite{AS}. 
 In fact, it can be applied whenever 
the free fermions energy momentum tensor can be written in term of a Sugawara form.  
Obviously, the light-cone momentum and Hamiltonian of  
any CFT that posses an affine Lie algebra and is coupled to non Abelian gauge field associated with the  
same algebra, can also be described in terms of holomorphic currents \cite{KS}.     
 
In this approach to ${\rm QCD}_2$, that has been studied in \cite{KS,AS,AFS3}, one constructs the Fock space of 
physical states by applying current creation operators on the vacuum of the system.   
The lowest  
physical states constructed by applying  current creation operators on the singlet vacuum, are those 
built from two currents. These type of states were introduced in \cite{AS}. 
In  \cite{AFS3} a detailed analysis of their spectra was done. This included writing down    
a 't Hooft like equation for the wave function of these  
two currents states  and  solving it  
   for the lowest massive state. 
  An excellent agreement with the DLCQ results \cite{GHK,AP,uwe}  was found. 
 In addition, the 't Hooft model and the large $N_f$ limit spectra were re-derived. 
 
However, it turns  out that one can also construct states by using only  one current creation operator  
when applied on the adjoint vacuum \cite{AS}.  The adjoint vacuum is  derived  
by acting on  the singlet vacuum with fermionic zero modes. In the case of adjoint fermions by a single  
adjoint zero mode and for fundamental fermions by quark anti-quark zero modes.  
In a scheme where only currents are being used one should be able to express the adjoint vacuum also in terms 
of currents.

The goal of the present paper is to currentize the adjoint vacuum,  
compute the Mass$^2$ of this state, to check whether there is a domain of  
parameters where it is an eigenstate and to compare it to the other massive states. 
This will be carried out in  massless QCD with fundamental quarks with   
$N_f$  flavors.

The outcome of our analysis is that in the large $N_f$ limit the one current state applied on the adjoint vacuum  
is indeed an eigenstate of the mass operator with a mass of $\sqrt{e^2 N_f\over \pi}$ 
. In the large $N_c$ limit this is not the case and in fact when acting with $M^2$ 
on this state, the dominant daughter state is a two current state. 
We have further analyzed the fermionic structure of these states,
especially in the large $N_c$ limit, to connect to the 't Hooft analysis \cite{thooft}. 
 
On route to these results we were facing two technical obstacles.  
The first is  the normalization of the states built by multi-current creation operators, and the second is IR divergences. In certain parts of the computations 
we were able to regularize the IR divergences, but in others we left it as an open problem. Assuming 
that a regularization scheme can be found, we have fully determined the normalizations of the relevant states.

The paper is organized as follows.  
In section 2 the general setup of the currentization of massless ${\rm QCD}_2$ is written down. This includes 
the fermionic and bosonized  
action, the light cone gauge fixing, the light-front quantization, the momentum and Hamiltonian operators  
expressed in terms of holomorphic currents as  
well as the algebra of the currents. Section 3 is devoted to the construction of the one current states. 
The adjoint vacuum is  derived by applying   a current derivative,  
at zero momentum, on the singlet vacuum. The momentum eigenvalue of the one current state is then the momentum of the applied current on the adjoint vacuum. 
We act with $M^2= 2P^+P^-$ on the one current states in section 4.  
It is shown that in general these states are not eigenstates of $M^2$. 
In the large $N_f$ limit they are. In the large $N_c$ limit the two current state is shown to be dominant.  
The description of the states in terms of their fermionic underlying structure is presented in section 5. 
We summarize the results and present a short discussion in section 
6.  In Appendix A we derive some useful formulas for computing traces 
over generators in the adjoint representation of $SU(N)$, which appear 
in the normalization of the two current state.

\section{ A brief review of massless ${\rm QCD}_2$ currentization on the light-front}

The theory of  ${\rm QCD}_2$ with  
 multi-flavor massless  fermions in the fundamental 
representation of $SU(N_c)$ 
 is described by the following action 
\beq 
\label{qcd} S=\int d^2x \ {\rm tr}\ (-{1\over 2e^2}F_{\mu\nu}^2+i\bar \Psi\Dslash\Psi ) 
 \eeq 
where $\Psi = \Psi ^i_a$, $i=1 \dots N_c$, $a=1 \dots N_f$ and $D_\mu = \partial_\mu + i A_\mu$. 
 
An alternative description is achieved by bosonizing the theory. The bosonization of  
multi-flavor massive ${\rm QCD}_2$ is complicated since one has to translate the fermions into bosonic variables which are group elements of  $U(N_F\times N_c)$ \cite{FS}. 
However, for massless fermions one can use bosonization in the  
$SU(N_c)\times SU(N_f)\times U_B(1)$ scheme. 
This scheme is quite useful since  it  decouples  
the color and flavor degrees of freedom.  
The bosonized form of the action in this scheme is given by \cite{DFS,FSreview} 
\bea 
\label{bosonized}  S_{b} &=& 
  N_f S_{WZW}(h) + N_c S_{WZW}(g)  + \int d^2x\ {1\over 2}\partial _\mu \phi \partial ^\mu 
 \phi -\int d^2x\ {\rm tr}\ {1\over {2e^2}} F 
 _{\mu \nu} F^{ \mu \nu}  
\nonumber \\ 
 && +{N_f\over 2\pi} \int d^2 x \ {\rm tr}\ (ih^\dagger \partial_+ h A_- 
+ih\partial_ - h^\dagger A_+ - A_+ h A_- h^\dagger + A_+ A_-)   \nonumber 
\eea 
where $h \in SU(N_c)$, $g\in SU(N_f)$, $\phi$ is the bosonic field for the baryon number 
 and $S_{WZW}$ stands for the 
Wess-Zumino-Witten action, which reads 
\ber 
\lefteqn{S_{WZW}(g)={1\over{8\pi}}\int _\Sigma d^2x \ {\rm tr}\ (\partial _\mu 
g\partial ^\mu g^{-1}) + } \\ 
 && {1\over{12\pi}}\int _B d^3y \epsilon ^{ijk} \ 
 {\rm tr}\ (g^{-1}\partial _i g) (g^{-1}\partial _j g)(g^{-1}\partial _k g), 
\eer 
 $B$ is a three dimensional volume whose boundary is the two dimensional surface $\Sigma$, which in our case is the 1+1 Minkowsky space, of one space and one time. 
 
As was advertised,  the flavored sectors are indeed decoupled from the colored one. 
Moreover the former ones are entirely massless. Hence,  
since we are interested in the massive spectrum of the theory  
 we will set  aside the $g$ and $\phi$ fields and analyze only the colored field $h$. 
In fact in \cite{KS} it is argued that there  
is a residual interaction of the zero modes of the $g,h$ and $\phi$ fields, but 
this will  not be important to our discussion. 
 
Next we choose the light cone gauge $A _- = 0$, and quantize the system on the  
light-front.  
Upon   integrating $A_+$ we 
find the following non local action   
\beq 
S = N_f S_{WZW}(h) - {1\over 2} e^2 \int d^2x\ {\rm tr}\  
({1\over \partial _-} 
J^+)^2 , 
\eeq 
where $J^+ = {i N_f\over 2\pi} h\partial _- h^{\dagger}$.  
The light-front momentum and Hamiltonian 
 take now simplified  forms. The momentum 
\beq  
P^+={1\over{N_c+N_f}}\int dx^- :J^a(x^-,x^+=0)J^a(x^-,x^+=0):  
\eeq 
namely, the Sugawara form , and the energy 
\beq 
P^-=-{e^2 \over {2\pi}}\int dx^-:J^a(x^-,x^+=0){1\over{\partial _- 
^2}}J^a(x^-,x^+=0):  
\eeq 
where  $J= \sqrt \pi J^+$. 
In the light-front the mass of any   state is given by the 
 eigenvalue equation 
\beq 
2P^+P^-\state{\psi} = M^2\state{\psi}. 
\eeq 
 
Three remarks are now in order: 
\begin{itemize} 
\item 
It is clear that we have indeed fully {\it currentized} the system in the  
sense that the equations that determines the spectrum are entirely expressed 
in terms of currents.   
\item 
Moreover both $P^+$ and $P^-$ depend only on $J^+$ and not on $J^-$ 
(obviously we could have a dependence only on $J^-$ by choosing the $A_+=0$ gauge 
and the space coordinate to be $x^+$). 
\item 
Furthermore, it is clear from the derivation of the above that the only condition 
for   currentizing  a theory of fermions coupled to non Abelian gauge fields, is that 
$T^{++}$ of the free fermionic theory could be rewritten in terms of a Sugawara form. 
This question was analyzed in \cite{GNO}. 
In particular in  
 \cite{AGSY,FSreview} it was shown that this can be done for adjoint fermions and 
in \cite{AFS} for  fermions in the symmetric and antisymmetric ``two box'' 
representations.    
 
\end{itemize} 
 
Next we write $P^+$ and $P^-$ in terms of the Fourier transform of $J(x^-)$ 
defined by $J(p^+)=\int {dx^- \over {\sqrt{2\pi}}} e^{-ip^+x^-} J(x^-,x^+=0)$. 
Normal ordering in the expression of $P^+$ and $P^-$ are naturally with respect 
to $p$, where $p<0$ denotes a creation operator. 
 To simplify the notation we 
write from now on $p$ instead of $p^+$. In terms of these variables 
the momenta 
generators are  
 
\begin{eqnarray} 
P^{+} & = & \frac{2}{N_{f}+N_{c}}\int_{0}^{\infty}dpJ^{b}(-p)J^{b}(p) 
\label{eq:definitionofp^{+}}\\ 
P^{-} & = & \frac{e^{2}}{\pi}\int_{0}^{\infty}dp\Phi(p)J^{b}(-p)J^{b}(p)\\ 
\Phi(p) & = & \frac{1}{2}\left(\frac{1}{(p+i\epsilon)^{2}}+ 
\frac{1}{(p-i\epsilon)^{2}}\right), 
\end{eqnarray} 
 $\epsilon$ is (as usual) a small parameter used to regulate the 
Fourier transform of $\int d^{2}xTr\left(\frac{1}{\partial_{-}}J^{+}\right)^{2}$. 
$\Phi(p)$ is $(-\frac{\sqrt\pi}{2})$ times the Fourier transform of the 'potential' $\left|x-y\right|$ 
between the currents. \footnote{Note the extra 2 factor  
as compared with eq.(7) of \cite{AFS3}, 
due to an error in the latter.} 
 
An important step in computing the eigenvalues of $M^2=2P^+P^-$ is interchanging the locations of  
current creation and annihilation operators. For this purpose we make use of the   
 level $N_f$, $SU(N_c)$ 
affine Lie algebra  that the  
 light-cone currents $J^a(p)$ obey   
\beq 
[J^a(p),J^b(p')]= {1\over 2}N_f\ p\ \delta ^{ab} \delta (p+p')+ 
 if^{abc} J^c(p+p'). 
\label{Kac} 
\eeq 
The Fock space of physical states is constructed as follows. First. the vacuum $\state{0,R}$ is defined 
as usual by the annihilation property  
\beq 
\forall p>0,\ J(p)\state{0,R}=0 
\eeq 
where R is an ``allowed"  representation depending on the level. 
 Physical states are built by applying the current creation operators on the vacuum, 
$\state{\psi} ={\rm tr}\ J(-p_1)\ldots J(-p_n)\state{0,R}$. Note that 
this basis is not orthogonal. 

\section{"The Adjoint Vacuum and its One-Current State"}
  
We would like to construct a state, which is obtained by the action of a  
current on the ``adjoint vacuum'', in the color singlet combination.  
  
This way we get a physical state, which is in a sense a ``one current'' state.  
Such states were considered before, in a different way, and the proof that   
they are really physical was not complete \cite{FHS,KS,AS}.  
    
The 'adjoint vacuum' is created from the singlet vacuum by applying 
the adjoint zero mode \cite{AS,AFS3}, which is taken as the limit $\epsilon\rightarrow0$ 
of the product of a quark and anti - quark creation operators, each 
one at momentum $\epsilon$ . hence in our case 
\begin{equation} 
\left|0,\; R\right\rangle =\lim_{\epsilon\rightarrow0}b_{\beta}^{\dagger i}(\epsilon) 
d_{j}^{\dagger\beta}(\epsilon)\left|0\right\rangle \label{eq:definitionofzm(limit)} 
\end{equation} 
where  $b_{\beta}^{\dagger i}$ and  $d_{j}^{\dagger\beta}$ are the creation operators of a quark and anti-quark respectively (see section 5). 
We can represent the action of the above adjoint zero mode on the 
vacuum by the derivative of a creation current taken at zero momentum. 
Differentiating  the current  with respect to $k$, and acting 
on the vacuum we get: 
\bea 
J_{j}^{'i}(k)\left|0\right\rangle _{k=0^{-}} & = &\sqrt{\frac{\pi}{2}} \frac{d}{dk}\int_{0}^{\infty}dp\int_{0}^{\infty}dq 
\delta(k+p+q) b_{\beta}^{\dagger i}(p)d_{j}^{\dagger\beta}(q)\left|0\right\rangle _{k=0^{-}} \nonumber \CR 
 & = &-\sqrt{\frac{\pi}{2}} b_{\beta}^{\dagger i}(\epsilon)d_{j}^{\dagger\beta}(\epsilon)\left|0\right\rangle _{\epsilon\rightarrow 0.} \nonumber \\ 
\eea 
As the currents are traceless, we have to subtract the trace part for $i=j$. 
The latter can be neglected for large $N_c$. 

For any given $N_c$, our results that follow are the same also after the trace 
is subtracted. 

The adjoint vacuum we have is a bosonic one, constructed from 
fermion-antifermion zero modes, and as we show can be written as the
derivative of the current acting on the singlet vacuum.
In the case of adjoint fermions there is another adjoint vacuum, 
a fermionic one, obtained by applying the adjoint fermion zero mode on 
the singlet vacuum.

As we showed already, $(J^a)^{'}(0)\left|0\right\rangle$ 
represents the adjoint zero mode  \\
$b^{\dagger}(0)d^{\dagger}(0)\left|0\right\rangle$ (indices suppressed),
for any $N_f$ and $N_c$,
so in particular also for $N_f=N_c$. But in the latter case the theory is
equivalent to that of adjoint fermions, as stated by the equivalence 
theorem in \cite{KS}. As also stated there, states built on the adjoint 
vacuum above, cannot be distinguished from those built on the fermionic 
adjoint vacuum, the latter obtained by applying the adjoint fermions on 
the singlet vacuum.

The adjoint bosonic vacuum can have also flavor quantum numbers, 
when the fermion have flavor. 
This does not change our results about the mass of the new state
we have. Our "currentball" will have flavor too in such a case.
In our scheme of bosonization, which is the "product scheme", especially 
convenient when the quarks are massless, the flavor sector is decoupled, and 
so the flavor multiplets are given by the action of flavor zero modes, not 
changing the mass values [see next Section].

Let us introduce the notation  
$$ Z^a \equiv -\sqrt{\frac{2}{\pi}} (J^a)^{'} (0).$$  
The state we have in mind is   
                $$|k\rangle = J^{b}(-k) Z^b |0\rangle.$$  
This state is obviously a global color singlet, but in our Light Cone gauge   
$A_-=0$ it is also a local color singlet, as the appropriate line integral   
vanishes.  
  
Now  
\be 
\sqrt{\frac{\pi}{2}}\left[J^{a}(p),Z^{b}\right]=\frac{1}{2}N_{f}\delta^{ab}\delta(p)-if^{abc}  
(J^{c})^{'}(p),
\ee  
and thus, for $p>0$  
  
$$  
J^{a}(p)Z^{b}|0\rangle=Z^{b}J^{a}(p)|0\rangle-i\sqrt{\frac{2}{\pi}}f^{abc}(J^{c})^{'}(p)|0\rangle=0.  
$$  
Hence the state $Z^{b}|0\rangle$ is annihilated by all the annihilation  
currents, and so it is indeed a colored vacuum.  
  
Using   
\be  
\left[P^{+},J^{b}(-k)\right]=kJ^{b}(-k)  
\ee  
we get that our state $|k\rangle$ is indeed of momentum $k$.  

Note that when quantizing on a circle of radius $R$, the adjoint vacuum
would be an eigenstate of $P^+$ with eigenvalue $N_c/R$ \cite{GW}.
As we work in the continuum limit, we get zero.

\section{The action of $M^2$ on the one current states} 
  
First, we evaluate the commutator of $P^{-}$ with a creation current  
  
  
\begin{eqnarray*}  
\lefteqn{\left[\int_{0}^{\infty}dp\phi(p)J^{a}(-p)J^{a}(p),J^{b}(-k)\right]  =  }\\  
 & &\frac{1}{2} N_{f} \frac{1}{k} J^{b}(-k) + if^{abc} \int_{0}^{k}{dp}\phi(p)J^{a}(-p)J^{c}(p-k)  
\\  
 && + if^{abc} 
 \int_{k}^{\infty}dp\left(\phi(p)-\phi(p-k)\right)J^{a}(-p)J^{c}(p-k)
\end{eqnarray*} 
note that in $P^{-}$ (and in $P^{+}$) we ignore contributions from  
zero - mode states, that is, we cut the integrals at $\epsilon$,  
and then take the limit.   
  
As $P^{+}$ and $P^{-}$ act on a singlet state, and as $J^a(0)$, being the
color charge, annihilates this state, the contribution from the zero
modes in both $P^+$ and $P^-$ is zero. Therefore it is legitimate to cut the
integration limit above the zero mode and then take the cutoff to zero,
as we have done. Note also that the integral of $\phi(p)$ around $p=0$ is
finite, and in fact zero when integrating over the whole line, therefore 
there are no divergences when we take the limit.

It is important, however, to remember that the zero mode does contribute
when we act upon non singlet states, like the adjoint vacuum $Z^b|0>$
itself. As mentioned above, when quantizing on a circle of radius R one
gets that $P^+$ is of order 1/R. And then, with $P^-$ of order $e^2 R$, 
$M^2$ is R independent, and so remains finite in the continuum limit. 
However, this is subtle, as $P^-$ becomes IR divergent in the continuum 
and needs to be regularized. This subtlety does not affect our calculation 
as we work in the singlet sector only.

Actually, the argument connected with $P^-$ acting on singlets should be
somewhat sharpened. Let us put the lower limit at $\epsilon$, and let it go to 
zero at the end. Then $J(\epsilon)$, when acting on a singlet, would go like 
$\epsilon$. We have two currents in the integral, so we get $\epsilon^2$.
But then we have $1/\epsilon^2$ from the denominator, so a finite
integrant. But the region of integration for is of order $\epsilon$, so indeed 
total contribution goes to zero.

Now apply $P^{-}$ on our state   
  
\be  
P^{-}J^{b}(-k)Z^{b}|0\rangle =   
 \left[P^{-},J^{b}(-k)\right]Z^{b}|0\rangle 
\ee  
as the Hamiltonian annihilates the color vacuum as well.  
  
Using the commutator of the Hamiltonian with a current, which we evaluated   
before, we get  
  
\begin{eqnarray*}  
\frac{\pi}{e^2} P^{-} J^{b}(-k) Z^{b}|0\rangle & = & \frac{1}{2} N_{f} \frac{1}{k} J^{b}(-k)Z^{b}|0\rangle\\ & + &   
if^{abc} \int_{0}^{k}dp\phi(p)J^{a}(-p)J^{c}(p-k)Z^{b}|0\rangle.  
\end{eqnarray*}  
Note that we used the fact that annihilation currents do annihilate also the   
colored vacuum.

Let us apply the operator $M^{2}$ to our one-current state  
   
$$ M^{2}J^{b}(-k)Z^{b}|0\rangle = 2P^{-}P^{+}J^{b}(-k)Z^{b}|0\rangle =   
 2k P^{-}J^{b}(-k)Z^{b}|0\rangle = $$  
\be  
(\frac{e^{2}N_{f}}{\pi})J^{b}(-k)Z^{b}|0\rangle+\\  
(\frac{2e^{2}}{\pi}k)if^{abc}\int_{0}^{k}dp\phi(p)J^{a}(-p)J^{c}(p-k)Z^{b}|0\rangle.  
\ee  
So it seems that, in the large $N_{f}$ limit, the state $J^{b}(-k)Z^{b}|0\rangle$  
is an (approximate) eigenstate, with eigenvalue $\frac{e^{2}N_{f}}{\pi}$.  
  
To see the exact dependence of the two terms in the equation above  
(the one and two current states) on $N_{f}$ and $N_{c}$, we should  
normalize them.   
The normalization of $J^{b}(-k)Z^{b}|0\rangle$ is  
\begin{eqnarray} 
\lefteqn{\left\langle 0\right|Z^{a}J^{a}(k)J^{b}(-k)Z^{b}\left|0\right\rangle =\left\langle 0\right|Z^{a}\left[J^{a}(k),J^{b}(-k)\right]Z^{b}\left|0\right\rangle =}\nonumber \\ 
&&\frac{1}{2}N_{f}k\delta(0)\left\langle 0\right|Z^{b}Z^{b}\left|0\right\rangle   + if^{abc}\left\langle 0\right|Z^{a}J^{c}(0)Z^{b}\left|0\right\rangle =\\ 
&&\frac{1}{2}N_{f}k\delta(0)\left\langle 0\right|Z^{b}Z^{b}\left|0\right\rangle   +  N_{c}\left\langle 0\right|Z^{b}Z^{b}\left|0\right\rangle \nonumber \end{eqnarray} 
The second term in the last line can be neglected compared to the first, as it   
is a constant to be compared with $\delta(0)$ [the space volume divided by   
$2 \pi$].  
  
Now  
$$\langle0|Z^{b}Z^{b}|0\rangle = (N_c^2 - 1) \langle0|Z^{1}Z^{1}|0\rangle $$  
and the factor $k \delta(0)$ is the normalization of a plane wave of momentum   
k. So the normalized state is, for $N_c >> 1$,  
\be  
\frac{1}{N_{c}\sqrt{\frac{1}{2}N_{f}}}J^{b}(-k)Z^{b}|0\rangle  
\ee  
relative to $\langle0|Z^{1}Z^{1}|0\rangle$.  
  
The normalization of the second term is more complicated.  
A lengthy but straightforward calculation gives  
\begin{eqnarray} 
\lefteqn{\left\| \left(if^{def} k 
\int_{0}^{k}dq\Phi(q)J^{d}(-q)J^{f}(q-k)\right)Z^{e}\left|0\right\rangle
\right\|^{2}=}  
 \CR 
 && N_{c}\left(N_{c}^{2}-1\right)\left(\frac{1}{2}N_{f}\right)^{2} 
k \delta(0)\left\langle 0\left|Z^{1}Z^{1}\right|0\right\rangle\times  \\ 
 && k\left( \int_{0}^{k}dpp(k-p)\Phi(p)\left(\Phi(p)-\Phi(k-p)\right) 
-\frac{N_c}{N_f}\int_{0}^{k}dp\Phi(p)\int_{0}^{k-p}dq q
\Phi(q)\right)
 \nonumber 
\eea 
We have written only the terms proportional to $\delta(0)$ as they 
are the dominant ones. Useful formulae for the evaluation, involving sums of  
products of structure functions of $SU(N)$, are given in the Appendix. 
 
The various momentum integrals (including the ones for the non dominant 
terms) are divergent for $\epsilon\rightarrow0$, thus they should 
be regulated. We leave this problem for now, and assume henceforth 
that they are regulated and finite. For simplicity the integrals  
(including the factor k) appearing 
in the two dominant terms will be notated $R_{1}$ and $- R_{2}$ in 
the following expressions. Note that we have $\frac{1}{\epsilon^2}$ and
$\frac{1}{\epsilon}$ divergences and also $\ln(\frac{k^2}{\epsilon^2})$.  
It seems that these are cancelled in $R_2$.

Define now the normalized states 
\be 
|S_1\rangle = C_1 \left(J^{b}(-k)Z^{b}\left|0\right\rangle \right) 
\ee 
 
\be 
|S_2\rangle = i C_2 k f^{abc}\int_{0}^{k}dp\Phi(p)J^{a}(-p)J^{c}(p-k)Z^{b} 
\left|0\right\rangle 
\ee 
where 
\be 
C_1= \frac{1}{N_c\sqrt{\frac{1}{2}N_f}}, \qquad  
C_2=\frac{ \frac {2}{N_f\sqrt{N_c^3}}}{\sqrt{R_1 + R_2\frac{N_c}{N_f}}}. 
\ee 
 
The mass eigenvalue equation of the normalized states is  
\be 
M^2 |S_1\rangle = 
\frac{e^2}{\pi}N_{f} |S_1\rangle + 
 \frac{e^2N_c}{\pi} \sqrt{2} 
\sqrt{R_1\frac{N_f}{N_c} + R_2} |S_2\rangle, 
\ee 
thus, we see that in the large flavor limit, our state $|S_1\rangle$ is  
an eigenstate with mass   
\be  
M= \sqrt {e^2 N_f \over {2 \pi}}.   
\ee  
In the large color limit, however, we actually get   
that the   
second term dominates by a factor of $N_c$. Moreover, while the first term goes  
to zero in the large $N_c$ limit, due to the factor of $e^2$, the second term  
survives in that limit.  
\section{Fermionic structure of the states} 
 
The analysis of the spectrum of $QCD_2$ in the fermionic formulation was addressed in \cite{GKMS,DK,kutasov,BDK,dalley}. 
Here we do not intend to perform the full process of determining the spectrum of the theory 
with fundamental fermions but rather only to find out the fermion structure of our state. 
To this end we write it as  
\beq
J^i_j (-k)  b^{\dagger j}_\beta(\epsilon) d^{\dagger \beta} _i ( \epsilon)\left|0\right\rangle  
\label{state}
\eeq 
for $\epsilon>0$ tending to zero and where  
\be 
J^i_j (-k) = \frac{1}{2\pi}\int_0^\infty dp[ b^{\dagger i}_\beta(p+k) 
b^{\beta} _j (p) +  
\theta(k-p) b^{\dagger i}_\beta(k-p) d^{\dagger \beta} _j (p) -  d^{\dagger \beta}_j(p+k) d^{i} _\beta (p)]. 
\ee 
So the 4-quarks part in (\ref{state})  has a coefficient which is independent of $N_c$. 
 
As for the 2-quark part, it involves an anti-commutator of creation with annihilation, yielding a state which is a combination of  
\be 
b^{\dagger j}_\beta(k) d^{\dagger \beta} _j (\epsilon)\  \ \ and \qquad  b^{\dagger j}_\beta(\epsilon) d^{\dagger \beta} _j (k) 
\ee 
with a coefficient that is proportional to $N_c$. 
Thus for large $N_c$, we have a quark-antiquark combination of momenta $( k,0)$ and   $(0, k)$. 
  
As 't Hooft found all meson states for large $N_c$, and each has a well defined momentum distribution \cite{thooft}, it is clear that our state is not a mass eigenstate  
of large $N_c$.  This is of course part of our explicit calculation in the previous section.  
 
\section{Discussion}

In this note we have continued the investigation of  two dimensional massless multi flavor $QCD$ in its 
``currentized'' form. The currentization formalism serves as a universal description \cite{KS}  of systems  
of matter fields coupled to non Abelian gauge fields. In particular fermions in various  representations 
and WZW actions at any level. The currentization also sets a common framework for the dual pairs of  fermionic  
theories and their bosonized partners ( or vice versa). The hope has been of course that this method will also be  
a useful tool  
to solve  those $QCD_2$ systems.  
 
The concrete question that has been addressed in this paper is the spectrum of states constructed by applying a single 
current creation operator on the adjoint vacuum. For that purpose we had to write down the adjoint vacuum in terms of  
currents. We proved that it is given in terms of the derivative with respect to $k$, at $k = 0$, of the current acting on the singlet 
vacuum. We then applied the operator $M^2$ on  one current states built on the adjoint vacuum.  
We found that in general, and in particular also in the large $N_c$, these states are not eigenstates of $M^2$. However, 
in the large $N_f$ limit they are eigenstates.

In \cite{AFS3}, where  the spectrum of  two current states on the singlet vacuum was derived, it was shown that  
in the large $N_f$ limit  there is a continuum of states with mass above $2\times e\sqrt{\frac {N_f}{\pi}}$. 
This indicates that there is a non-interacting  ``currentball'' meson of mass $ e\sqrt{\frac {N_f}{\pi}}$. 
In our present work, when invoking  the large $N_f$ limit we have indeed found an eigenstate of $M^2$ 
with exactly  
this mass.
The state we have found is a color singlet. In fact it is easy to see that in the   large $N_f$ limit, there
are   $N_c^2-1$ colored eigensates of $M^2$ with the same mass. This can be interpreted as follows.  
In the large $N_f$ $QCD_2$ is transfered into a set of $N_c^2-1$ Abelian systems. This can be easily seen   
in (\ref{Kac}) where in the large $N_f$ limit the non Abelian term of the commutator is sub-leading.   
Now in  $QED_2$  it is well known that  the Schwinger mechanism  yields a massive state of mass $\frac{e}{\sqrt{\pi}}$. 
This Schwinger state is often considered as a bound state of an electron-positron.  
In the large $N_f$ the $M^2$ eigenstates are therefore just the Schwinger states appearing in a multiplicity of $N_c^2-1$.

For any finite $N_f$ only the singlet state remains in the spectrum, of course.
However, at least formally, in the infinite $N_f$ limit, we get that the
current algebra is like of $N_c^2 - 1$ currents, each one a U(1).

In \cite{FHS}, using a different currentization approach to $QCD_2$, it was conjectured that there should be 
also non-Abelian Schwinger mechanism that will yield massive physical states in the  color ``adjoint'' representation, 
the analogs of the massive particle of the Abelian Schwinger model.  
Since we have detected such eigenstates of $M^2$ in the large $N_f$ limit, and we found that without this limit they are 
not eigenstates, we conclude that if the non-Abelian Schwinger particles indeed exist, they mix with other states  
constructed by applying more than one current creation  operator on a vacuum 
state. 
 
Note, that when having adjoint fermions, there is also a fermionic adjoint 
vacuum, as we mentioned already in Section 3.
Trittmann \cite{UT} discussed this case.

There are certain open questions that follow from the analysis of this paper: 
\begin{itemize} 
 
\item 
As mentioned in section 4 the computation of the spectrum of the states requires a regularization prescription 
that we have only partially found.  
\item 
The ``ultimate'' goal  of this reseach work is to diagonalize the ``currentball'' states created  by applying any number of  
current creation operators on  the various vacua of the theory.  
\item  
On top of everything stands the question what have we learned from the currentization procedure and from the  
two dimensional spectrum of states about four dimensional $QCD$. In particular a challenging question is to investigate the 
possibility of a Schwinger like mechanism also in four dimensions.  
\end{itemize} 
 
{\bf Acknowledgments}  
 
The work of A.A. and J.S.  was supported in part by the Israel  Science Foundation and by the German Israeli Foundation. 
J.S. would like to thank the Einstein Center of Theoretical Physics at the Weizmann Institute for support.  
\section{Appendix} 
 
The generators $T^a$ of $SU(N)$, in the adjoint representation, are related to the structure constants $f^{abc}$ as  
$$ (T^a)_{bc} = -i f^{abc}.$$ 
Thus  
$$ f^{abc} f^{abd} = Tr (T^c T^d) = N \delta^{cd}$$ 
and  
\begin{eqnarray*}
\lefteqn{ f^{abc} f^{a'bc'} f^{aa'd} f^{cc'd} 
= Tr (T^b T^d T^b T^d) =} 
\CR && i f^{bde} Tr (T^e T^b T^d) + Tr (T^b T^b T^d T^d)  
= {1\over 2} N^2 (N^2 -1).
\end{eqnarray*}
where we used 
$$\sum_a T^a T^a = N I_{adj}$$ 
with $I_{adj}$ the unit matrix in the adjoint representation. 
 
 
\newpage

\appendix

\end{document}